\documentclass{aa}

\usepackage{graphicx}

\usepackage{txfonts}
\usepackage{color}

\begin{document}

\title{Dynamical analysis of the complex radio structure in 3C\,293:\\ Clues on a rapid jet realignment in X-shaped radio galaxies}

\titlerunning{Dynamical analysis of the radio structure in 3C\,293}

\author{J. Machalski\inst{1}, M. Jamrozy\inst{1}, {\L}. Stawarz\inst{1}, \and M. We\.{z}gowiec\inst{1}}

\authorrunning{Machalski et al.}

\institute{Astronomical Observatory, Jagellonian University, ul. Orla 171, PL-30244 Krakow, Poland \\ \email{machalsk@oa.uj.edu.pl}}

\date{}

\abstract{Radio galaxies classified as X-shaped/winged, are characterised by two pairs of extended and misaligned lobes, which suggest a rapid realignment of the jet axis, for which a potential cause (including binary supermassive black holes, a black hole merger, or a Lense-Thirring precession) is still under debate.}{Here we analyse the complex radio structure of 3C\,293 winged source hosted by the post-merger galaxy UGC\,8782, which uniquely displays a significant asymmetry between the sizes (and therefore the ages) of the two pairs of lobes, indicating that an episode of jet realignment took place only very recently. This allows us to tightly constrain  the corresponding timescales, and therefore to discriminate between different models proposed for the formation of X-shaped radio galaxies in general.}{Based on all the available and carefully re-analysed radio data for 3C\,293, we have performed a detailed spectral modelling for the older and younger lobes in the system, using the existing evolutionary DYNAGE algorithm. In this way we derived the lobes' ages and jet energetics, which we then compared to the accretion power in the source.}{We found that the $200$\,kpc-scale outer lobes of 3C\,293 are $\sim 60$\,Myr old and, until very recently, have been supplied with fresh electrons and magnetic field by the jets, i.e.,   jet activity related to the formation of the outer lobes  ceased within the last Myr. Meanwhile, the inner 4\,kpc-scale lobes, tilted by $\sim 40\degr$ with respect to the outer ones, are only about $\sim 0.3$\,Myr old. Interestingly, the best model fits also return  identical values of the jet power supplying the outer and the inner structures. This power, moreover, is of the order of the maximum kinetic luminosity of a Blandford-Znajek jet for a given black hole mass and accretion rate, but only in the case of relatively low values of a black hole spin, $a\sim 0.2$.}{The derived jet energetics and timescales, along with the presence of two optical nuclei in UGC\,8782, all provide a strong support to the Lense-Thirring precession model in which the supermassive black hole spin, and therefore the jet axis, flips rapidly owing to the interactions with the tilted accretion disk in a new tidal interaction episode of the merging process. We further speculate that, in general, X-shape radio morphology forms in post-merger systems that are rich in cold molecular gas, and only host  slowly spinning supermassive black holes.}


\maketitle

\section{Introduction}

The nearby ($z=0.0452$) galaxy UGC\,8782 with its complex bulge-dominated morphology, a well-defined gaseous disk with weak spiral arms, and prominent dust lanes (e.g., de Koff et al. 2000; Floyd et al. 2006), is very likely the result of a merger event (Martel et al. 1999), and probably a gas-rich merger of an elliptical galaxy and a spiral galaxy (Singh et al. 2015). This association is reinforced by the detailed stellar population analysis made by Tadhunter et al. (2005), which reveals a mix of an old stellar population and a relatively young stellar population in the system. In fact, {\sl Hubble} Space Telescope (HST) optical and UV images have disclosed several regions of a recent starburst formation (Baldi \& Capetti 2008). Multiwavelength imaging of UGC\,8782 shows an optical jet or a tidal tail extending towards a small companion galaxy about 30 kpc away in the south-west direction (Evans et al. 1999). For a further description of this interacting system, see Lanz et al. (2015).

Formerly, the large-scale radio structure of 3C\,293 hosted by the galaxy UGC\,8782 was classified as a Fanaroff-Riley\,I/II type with a bright steep-spectrum ``core'' (Argue, Riley \& Pooley 1978; Bridle, Fomalont \& Cornwell 1981). More recently, this core region was resolved with the MERLIN array (Akujor et al. 1996; Beswick, Pedlar \& Holloway 2002), which revealed a compact ($\sim 2$\,arcsec) pair of symmetric and steep-spectrum lobes supplied by small-scale jets, and extending in an east-west direction, with adjacent less luminous tail-like feature bending out of the galactic disk. The angular size of this feature is about $\gtrsim 4$\,arcsec, i.e., about 4\,kpc in projection. The complex jet morphology of the inner 2-arcsec structure and the position of the steeply inverted-spectrum unresolved radio nucleus are discussed by Beswick et al. (2004), based on a combination of the VLA, MERLIN, and global VLBI network observations.

The large-scale radio continuum emission of 3C\,293, i.e., the outer lobes with the projected total size of about 3.6\,arcmin, or 190\,kpc, has been mapped in detail by Beswick et al. (2004). There is a conspicuous difference between the space orientation of the inner 4\,kpc-scale structure and the outer lobes extending at a position angle $\sim 40\degr$. There is also some evidence for the large-scale jet contribution to the emission of the extended outer radio lobes, which could be seen on the 1465\,MHz total intensity and linearly polarized intensity VLA maps (resolution of 6 arcsec) presented by Bridle et al. (1981). A clearly visible bridge emission extends northwest from the core and terminates in the extended lobe some 75\,kpc from the core. The structure is brighter close to the nucleus and connects directly with it, rather than fading near the nucleus, as in other classical doubles. As suggested by Bridle et al. (1981), the complicated oscillatory internal structure of the spine of this bridge and its high degree of linear polarization (up to 40\%) resembles a disrupted jet structure.

In this paper, we consider and analyse a possibility that the outer and inner lobes of 3C\,293, forming together the X-shaped structure, result from two distinct episodes of the jet activity in the source: the older one relates to an early phase of the merging process, while another one was only recently triggered/reshaped by a new tidal interaction episode. The above would correspond to a conclusion made by Tadhunter et al. (2011) in their study of evolutionary histories and triggering in starburst radio galaxies: ``... majority of the starburst radio galaxies have been triggered in galaxy mergers in which at least one of the galaxies is gas rich. However, the triggering (or re-triggering) of the radio jets can occur immediately before, around, or a significant period after the final coalescence of the merging nuclei, reflecting the complex gas infall histories of the merger events.'' Indeed, the archival HST image taken with the WFPC2 facility, reproduced here in Figure\,1, shows the central part of the galaxy with two optical nuclei apparently straddling the central dust lane along the east-west axis (see Martel et al. 1999). The separation of the nuclei is $\sim 1$\,arcsecs\,$\sim 880$\,pc only. The western nucleus strictly coincides with the unresolved radio core.

We note that 3C\,293 is not a classical X-shaped radio source according to the standard definition adopted in Cheung (2007), who compiled a large sample of these types of objects using the VLA FIRST survey database. In particular, the radio galaxy analysed here does not contain a pair of misaligned, low surface brightness but does containt extended `wings', in addition to the primary pair of FR\,II-type lobes. However, Dennett-Thorpe et al. (2002) have compared 3C\,293 to the archetype X-shaped radio galaxies, concluding: ``Its structure could be interpreted as an extreme case of winged source with very short active lobes''. Interestingly, the total radio power of the source, $P_{\rm 1.4\,GHz} \simeq 2.2\times 10^{25}$\,W\,Hz$^{-1}$, and its narrow line region (NLR) luminosity, $L_{\rm NLR} \sim 1.5 \times 10^{34}$\,W, are both intermediate between the FR\,I and FR\,II classes, and characteristic for so-called classical X-shaped radio galaxies in general (Landt et al. 2010). In the above, we  estimate the NLR luminosity based on the observed line luminosities $L_{\rm [O\,{\sc II}]} \simeq L_{\rm [O\,{\sc III}]} \simeq 10^{33}$\,W (Emonts et al. 2005, Sikora et al. 2013), following the scaling relations of Rawlings \& Saunders (1991) and Willott et al. (1999). Furthermore, the emission-line gas density in the system, determined from the flux ratio of $1.2 \pm 0.1$ between the sulphur line doublet S{\sc [II]}$\lambda$6716/$\lambda$6731 (Emonts et al. 2005), is relatively low, which is again in agreement with  what has been established for the FIRST sample of X-shaped radio sources (Landt et al. 2010).

To perform a detailed and robust dynamical modelling for different components of the observed complex radio structure of 3C\,293, all the corresponding radio spectra and linear sizes must be determined precisely. For this reason, in Section~2, we present the archival data and radio maps used in our analysis. The analysis is initially focused upon the extraction of the emission of the large-scale radio lobes from the emission of the entire source when observed with poor angular resolution, especially at long wavelengths. Subsequently, we attempt to separate the emission of the inner lobes from the radiative output of small-scale jets. The dynamical ageing analysis and the best-fit models for the outer and the inner radio lobes are described in Section~3, and the auxiliary results of the {\sl Chandra} X-ray observations of the system are summarized in Section~4. Further discussion regarding the emerging constraints on the jet energetics and duty cycle, in the general context of formation of X-shaped radio galaxies, is given in Section~5.

In this paper, we assume a modern flat cosmology with $\Omega _{\rm m}=0.27$, $\Omega_{\Lambda}=0.73$, and $H_{0}=71$\,km\,s$^{-1}$\,Mpc$^{-1}$, so that the source redshift of $z=0.0452$ corresponds to the luminosity distance of $d_{\rm L} = 197.6$\,Mpc and the conversion scale of $0.877$\,kpc/arcsec.

\begin{figure}[t]

\centering

\includegraphics[width=90mm,angle=0]{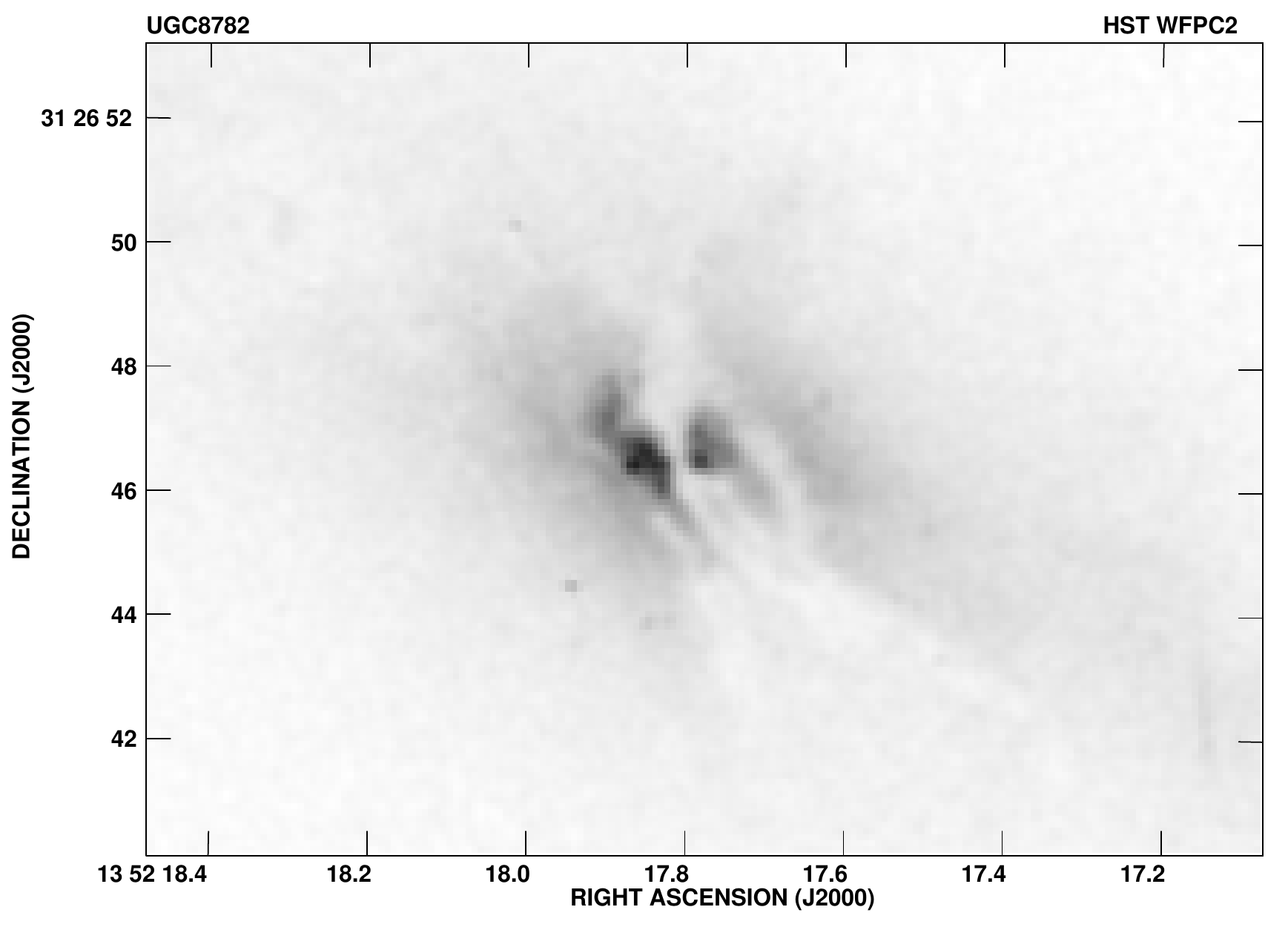}

\caption{HST WFPC2 $R$-band negative image of the central part of the galaxy UGC\,8782, showing the bulge region with prominent dust lanes and two optical nuclei (see a positive image in de Koff et al. 2000).}

\end{figure}

\section{Radio maps and data} 

The radio maps used to determine the sizes and monochromatic luminosities of the outer and the inner lobes in 3C\,293, are presented in Figure\,2 and Figure\,3, respectively. 
 These images (except Fig.~3(d), which is reproduced directly from Beswick, Pedlar \& Holloway (2002)), are obtained reducing the archival VLA data in a standard manner using the AIPS package.

Figure\,2 shows the 1.36\,GHz VLA B-array (project code GP022) image of the outer lobes (this map is similar to that shown by Beswick et al. 2004 in their Fig.\,2). Figure\,3 shows four images of the inner structure at higher observing frequencies, i.e., at 1.36\,GHz (MERLIN), 4.86\,GHz (VLA project code AT249), 14.94\,GHz (VLA project code AM067), and 22.46\,GHz (VLA project code AA149). The archival flux densities used to determine the radio spectrum of the entire source, as well as those precisely measured for the spectrum of the inner structure, are given in Table~1. The corresponding spectra are shown in Figure\,4a; since the flux densities of both components are comparable, the spectrum of the inner structure is shifted in the figure one-decade down, for clarity (cf. the flux density scale on the right-hand frame of the figure).

\begin{figure}[t]

\centering

\includegraphics[width=90mm,angle=0]{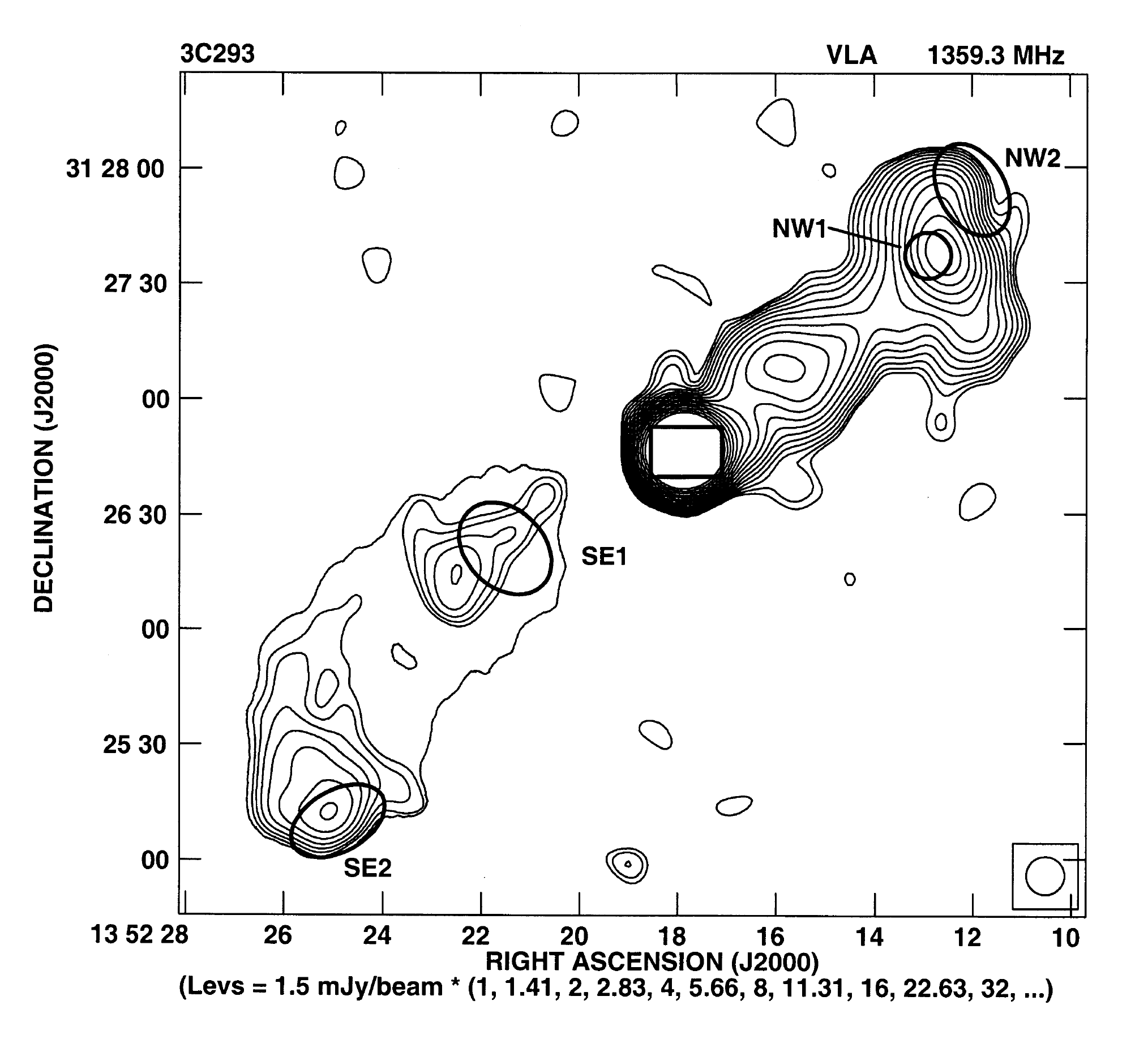}

\caption{VLA B-array 1.36\,GHz map of the large-scale structure of the radio source 3C\,293, hosted by the galaxy UGC\,8782. The rectangle inside the central core region corresponds to the frame of Figure\,1. The small circle and three ellipses denote the regions where the X-ray emission was detected and modelled by Lanz et al. (2015). The restoring beam ($10\arcsec\times10\arcsec$) is indicated by a circle in the bottom right corner.}

\end{figure}

\subsection{Outer lobes}

To extract the emission of the large-scale outer lobes from the radiative output of the entire source, we smooth out the radio spectrum of the inner structure, fitting the flux densities given in column\,5 of Table\,1 with the analytic formulae $y=A+Bx+C\exp(Dx)$, where $y=\log F_{\nu}$ is the logarithm of a flux spectral density, $x=\log \nu$ is the logarithm of the observed frequency, and $D=\pm 1$. Because both fits (corresponding to the two different values of $D$) differ  by only about 1\% to 3\% at the highest and the lowest observing frequencies, respectively, we used their mean values to extrapolate the spectrum of the inner structure to the frequency of 22 MHz. The resulting flux densities are given in column\,7 of Table\,1 and these are used to determine the correct spectrum of the outer lobes, except for the fluxes at  frequencies of 154, 240, 614, and 4860 MHz, which we adopt directly from the GMRT measurements by Joshi et al. (2011). The final spectrum of the outer lobes, obtained by subtracting the above values from the total flux densities of the entire source given in column\,2 of Table\,1, is shown in Figure\,4b, and the corresponding flux densities are given in column\,2 of Table\,2.

\begin{table}[t!]

\scriptsize{

\begin{center}

\caption{Archival flux densities of the entire 3C\,293 source, and of its inner structure, as well as a smooth fit to the flux densities of the inner structure (see Sect.\,2.1).}

\vspace{2mm}


\begin{tabular*}{87mm} {@{}rr@{}c@{}lr@{}c@{}lr@{}r@{}lr@{}c@{}lr@{}c@{}lc@{}c}

\hline

\hline

\multicolumn{1}{c}{Freq.} &\multicolumn{3}{c}{Entire source} &\multicolumn{3}{c}{Ref.} &\multicolumn{3}{c}{Freq.}

&\multicolumn{3}{c}{Inner structure} &\multicolumn{3}{c}{Ref.} & Smooth fit \\

\multicolumn{1}{c}{[MHz]} &\multicolumn{3}{c}{[mJy]} &\multicolumn{3}{c}{     }&\multicolumn{3}{c}{[MHz]}

&\multicolumn{3}{c}{[mJy]} &\multicolumn{3}{c}{    }& [mJy] \\

\multicolumn{1}{c}{(1)} &\multicolumn{3}{c}{(2)} &\multicolumn{3}{c}{(3)}&\multicolumn{3}{c}{(4)}

&\multicolumn{3}{c}{(5)} &\multicolumn{3}{c}{(6)}& (7) \\

\hline

\\

  22 & 66000 & $\pm$&8000& (12)& & & &     & &       &      &    & & & & 48245 \\

  38 & 49560 & $\pm$&5000& (10)& & & &     & &       &      &    & & & & 35877 \\

  80 & 31800 & $\pm$&5000& (9) & & & &     & &       &      &    & & & & 23370 \\

  86 & 27400 & $\pm$&1900& (10)& & & &     & &       &      &    & & & & 22397 \\

     &       &      &    &     & & & & 151 & & 16080 & $\pm$&1800&(14) & & & 15884 \\

 154 & 22288 & $\pm$&3340& (8) & & & & 154 & & 15570 & $\pm$&1600& (8) & & & 15691 \\

 160 & 18500 & $\pm$&2850& (13)& & & & 178 & & 14100 & $\pm$&1100& (9) & & & 15322 \\

 240 & 15142 & $\pm$&2270& (8) & & & & 240 & & 11041 & $\pm$&1656& (8) & & & 11864 \\

 318 & 11550 & $\pm$&900&  (9) & & & &     & &       &      &    &    & & &  9918 \\

 325 & 14646 & $\pm$&600&  (11)& & & &     & &       &      &    &    & & &  9762 \\

 365 & 11207 & $\pm$&277&  (6) & & & &     & &       &      &    &    & & &  9055\\

 408 & 11070 & $\pm$&850&  (9) & & & & 408 & &  8900 & $\pm$&900& (1) & & &  8418 \\

 614 &  8282 & $\pm$&587&  (8) & & & & 614 & &  6446 & $\pm$&451& (8) & & &  6426 \\

 750 &  7300 & $\pm$&225&  (10)& & & & 750 & &       &      &   &     & & &  5620 \\

 750 &  7630 & $\pm$&130&  (9) \\

 1400 & 4849 & $\pm$&80&   (4) & & & & 1407 & & 3750 & $\pm$&300& (1) & & &  3681 \\

 1400 & 4558 & $\pm$&113& (10) & & & & 1465 & & 3610 & $\pm$&80&  (3) & & &  3566 \\

 1400 & 4684 & $\pm$&90&   (5) & & & & 1490 & & 3503 & $\pm$&75& (15) & & &  3518 \\

      &      &      &  &       & & & & 1665 & & 3190 & $\pm$&90& (15) & & &  3260 \\ 

 2695 & 2922 & $\pm$&150& (10) & & & & 2695 & & 2400 & $\pm$&80&  (2) & & &  2351 \\

 4850 & 1840 & $\pm$&163&  (7) & & & & 4860 & & 1490 & $\pm$&70& (15) & & &  1501 \\

 4860 & 1909 & $\pm$&135&  (8) & & & & 4860 & & 1553 & $\pm$&78&  (8) & & &  1501 \\

 5000 & 1860 & $\pm$&90&   (9) & & & & 4995 & & 1360 & $\pm$&60&  (1) & & &  1474 \\

10695 & 1040 & $\pm$&40&   (9) & & & & 8085 & & 1030 & $\pm$&45&  (2) & & &  1027 \\

10700 &  976 & $\pm$&56&  (10) & & & &10700 & &      &      &  &      & & &   826 \\

14900 &  740 & $\pm$&40&  (10) & & & &14940 & &  567 & $\pm$&60& (15) & & &   628 \\

15035 &  750 & $\pm$&50&   (3) & & & &15000 & &  550 & $\pm$&150& (1) & & &   626 \\

      &      &      &  &       & & & &22460 & &  384 & $\pm$&40& (15) \\      

      \\

\hline

\end{tabular*}

\end{center}

{\sl References:} (1) Argue, Riley \& Pooley 1996; (2) Bridle \& Fomalont 1978; 

(3) Bridle, Fomalont \& Cornwell 1981; (4) Condon, Cotton \& Broderick 2002;

(5) Croft et al. 2010; (6) Douglas et al. 1996; (7) Gregory et al. 1996;

(8) Joshi et al. 2011; (9) K\"{u}hr et al. 1981; (10) Laing \& Peacock 1980;

(11) Rengelink et al. 1997; (12) Roger, Costain \& Steward 1986;

(13) Slee 1995: (14) Waldram et al. 1996; (15) this paper.

}

\end{table}

\subsection{Inner lobes}

The observed emission of the inner ($\sim 4$\,kpc) lobes is likely contaminated by the small-scale jets. Fortunately, milli-arcsec observations with the MERLIN, VLA, and global VLBI networks (Beswick et al. 2004), enable us to trace the complex jet structure inside the diffuse emission component that is clearly visible on several radio maps of this region. Based on the Beswick et al.'s data sets for the jet components, namely their flux densities measured at 1.36\,GHz and 4.55\,GHz, we found  that the jet emission constitutes a significant amount of the total emission detected from the inner structure, namely $\sim 52\%$ and $\sim 55\%$ at these two frequencies. However, for the dynamical ageing analysis, monochromatic luminosities of the diffuse emission (i.e., of the lobes), free of  jet contamination, are needed. To estimate these luminosities, we identify  Beswick et al.'s features E1, E2, W1, and W2 as the jet knots, and evaluate the corresponding spectral index $\alpha_{4.5}^{1.3}$ of the inner jet emission as 0.685. Assuming that the slope of the jet spectrum below 1.3\,GHz should not be much flatter than $\alpha_{4.5}^{1.3}$, we take the single spectral index of 0.67 between the observed frequencies of 1.4\,GHz and 22\,MHz. In the high frequency range the situation is less certain. From the data provided by Bridle et al. (1981) and Akujor et al. (1996), we estimate that the jet flux density at 15\,GHz should not be higher than 300\,mJy, corresponding to the high frequency spectral index of 0.85, which we anticipate below. Thus estimated jet flux densities, given in column\,5 of Table\,2, are then subtracted from the entries in column\,5 of Table\,1, which provides clean flux densities of the inner lobes, given in column\,6 of Table\,2. The corresponding spectrum of the inner lobes is shown in Figure\,4b.

\section{Dynamical modeling}

\subsection{Numerical code and the fitting procedure}

The modelling presented below is based on the dynamical approach of Kaiser, Dennett-Thorpe \& Alexander (1997) for the evolution of lobe-dominated FR\,II type radio sources (hereafter the KDA model). This approach enables  an evaluation of the lobes' extension and radio luminosity for a given set of input physical parameters, including the jet power $Q_{\rm j}$, the lobes' lifetime $t_{\rm lb}$, the power-law slope of the electron energy distribution injected by the jets into the lobes $p = 1+ 2 \alpha_{\rm inj}$, where $\alpha_{\rm inj}$ is the injection spectral index, and the central density of the ambient medium $\rho_0$, simultaneously accounting for the cumulative effects of energy losses as a result of the adiabatic expansion of the lobes, synchrotron radiation, and inverse Compton scattering of the cosmic microwave background (CMB) photons. We note that, for the ambient medium at $>1$\,kpc distances from the active nucleus, we anticipate the standard King profile with a given core radius $a_{0}$ and a slope $\beta$.

\begin{figure*}[t]

\centering

\includegraphics[width=180mm,angle=0]{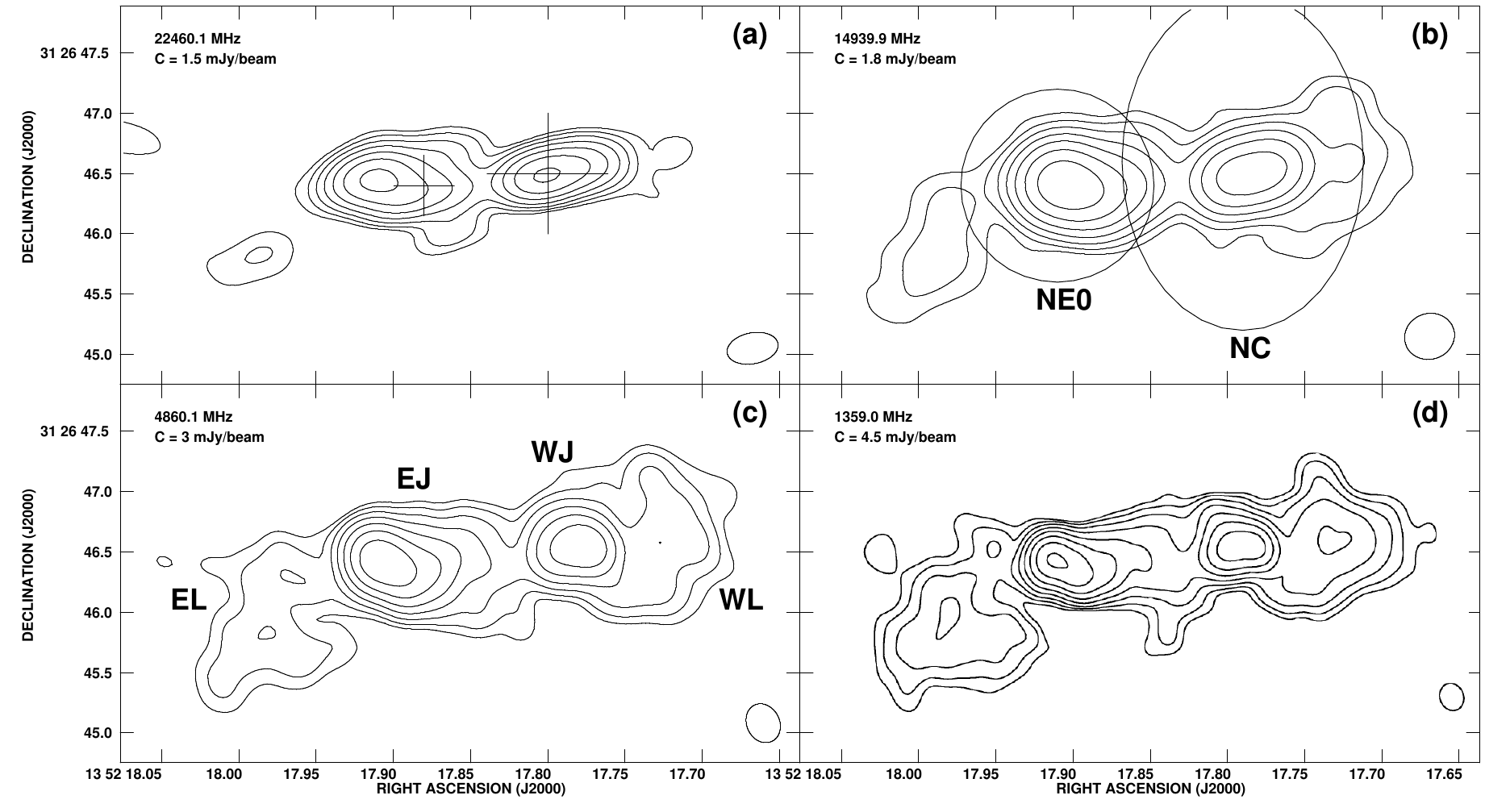}

\caption{Radio images of the inner radio structure of  3C\,293 at four different observing frequencies: (a) VLA B-array 22.46\,GHz; the large cross in the image indicates the position of the western optical nucleus/the active radio core, while the small cross denotes the position of the eastern optical nucleus; (b) VLA B-array 14.94\,GHz; the ellipse and the circle indicate the two smaller apertures selected by Lanz et al. (2015) for the {\sl Chandra} X-ray data analysis; (c) VLA A-array 4.86\,GHz; the labelling denotes the western lobe (WL), the western jet (WJ), the eastern lobe (EL), and the eastern jet (EJ); (d) MERLIN 1.36\,GHz reproduced from Beswick, Pedlar \& Holloway (2002). In all  panels, the contour levels given are (1, 2, 4, 8, ...)$\times$\,$C$\,mJy beam$^{-1}$, where the value of the $C$ constant are provided in each image. For all  images, the restoring beam is indicated by an ellipse in the bottom right corner.}

\end{figure*}

As in our previous works (Machalski et al. 2010, 2011), the modelling is performed using the DYNAGE algorithm of Machalski et al. (2007). This numerical code enables us to solve the inverse problem, i.e., to determine the above four free parameters ($Q_{\rm j}$, $t_{\rm lb}$, $p$, and $\rho_0$) by fitting the model to a given set of observables, including the linear size and the volume of the lobes, and the observed lobes' radio spectrum (i.e., the spectral power density at different observing frequencies). The summary of various model input parameters assumed in the modelling is given in Table\,3. We note that the values set for the viewing angle with respect to the jet axis, $\theta$, as well as the exponent of the ambient density distribution, $\beta$, are imposed observationally, by utilising the lobe/counterlobe radio brightness asymmetries, and the results of the {\sl Chandra} X-ray observations of the system, respectively.

\subsection{The modelling}

The evolution of the radio structure is modelled independently for the outer and the inner lobes. The results of the modelling are given in Table\,4 for three out of several solutions around the best-fit solution (shown in boldface). Subsequent lines in the table give some derivative physical parameters resulting from the modelling, including the longitudinal expansion velocity of the lobes, $\upsilon_{\rm h}$, the corresponding Mach number $\cal{M}_{\rm lb}$, the non-thermal pressure within the lobes, $p_{\rm lb}$, and the equipartition magnetic field strength, $B$.

A cursory inspection of the outer lobes' radio continuum in Figure\,3b, and of the corresponding data given in column\,2 of Table\,2, reveals that the difference between the low-frequency and the high-frequency slopes of the spectrum is less than 0.5. This implies no significant ageing (up to the observed frequencies of 15\,GHz) characteristic for dying/relict radio sources (see Kaiser \& Cotter 2002; Murgia et al. 2011). Indeed, following the fitting procedure described in Machalski et al. (2011), we find out that the best-fit model for the outer lobes of 3C\,293 is concordant with the  so-called continuum injection scenario, which means that the supply of fresh electrons and magnetic field by the jets is still ongoing; that is, the jet activity related to the formation of the outer lobes might have been terminated only very recently, with the upper bound given by the lobes' sound-crossing timescale $\tau_{\rm s/lb} = \sqrt{3} D /2 c \sin\theta \sim 0.7$\,Myr, assuming the sound speed within the lobes filled exclusively with ultrarelativistic gas $c_{\rm s/lb} = c/\sqrt{3}$. This, together with the estimated lifetime of the inner lobes of $\sim 0.3$\,Myr, constrains the timescale of a rapid realignment of the jet axis in 3C\,293. However, at the same time, the total jet kinetic power, $Q_{\rm j} \simeq 2 \times 10^{36}$\,W, did not change after the realignment episode.

We emphasise that the complex structure of the analysed radio galaxy differs substantially from that observed in the so-called double-double radio galaxies (DDRGs), which constitute yet another class of radio sources with clearly intermittent/recurrent jet activity. Generally, DDRGs are characterised by a co-linear pair of lobes stretching in the same direction, where inner and younger lobes are much  slimmer compared with the outer and older ones (e.g., Lara et al. 1999; Schoenmakers et al. 2000; Saripalli et al. 2002; Brocksopp et al. 2011). The DYNAGE analysis of outer and inner lobes of DDRG J1548--3216 presented in Machalski et al. (2010) reveals a significantly aged radio continuum of the outer lobes, and the jet power during the restarted activity about ten-fold fainter than that of the original jet. This implies very different physical conditions and processes responsible for re-triggering the jet activity in DDRGs and X-shaped radio galaxies. In this context the evolutionary scenario for classical doubles drafted by Liu (2004), in which a post-merger FR\,II system with binary black holes develops first an X-shaped morphology, and next, after the black hole coalescence, evolves into a DDRG structure.

\begin{table}[t!]


\scriptsize{

\begin{center}

\caption{Extracted flux densities and the best-fit flux densities for the outer and inner lobes of 3C\,293.}

\vspace{2mm}  

\begin{tabular*}{90mm} {@{}r r@{}c@{}l r@{}c@{}lr c r@{}c@{}lr@{}c@{}lc@{}c}

\hline

\hline

\multicolumn{1}{c}{Freq.} &\multicolumn{3}{c}{Outer lobes} &\multicolumn{3}{c}{Model fit} &\multicolumn{1}{c}{Freq.}

&\multicolumn{1}{c}{Jet est.} &\multicolumn{3}{c}{Inner lobes} &\multicolumn{2}{c}{Model fit} \\

\multicolumn{1}{c}{[MHz]} &\multicolumn{3}{c}{[mJy]} &\multicolumn{3}{c}{[mJy]}&\multicolumn{1}{c}{[MHz]}

&\multicolumn{1}{c}{[mJy]} &\multicolumn{3}{c}{[mJy]}&\multicolumn{1}{c}{[mJy]} \\

\multicolumn{1}{c}{(1)} &\multicolumn{3}{c}{(2)} &\multicolumn{3}{c}{(3)}&\multicolumn{1}{c}{(4)}

&\multicolumn{1}{c}{(5)} &\multicolumn{3}{c}{(6)} &\multicolumn{1}{c}{(7)}\\

\hline

\\

  22 & 17755 & $\pm$&2800 & {\sl 19264} \\

  38 & 13683 & $\pm$&1810 & {\sl 13619} \\

  80 &  8430 & $\pm$&1750 & {\sl  8398} \\

     &       &      &     &       & & & 151 & 8438 & 7642&$\pm$&1705 & {\sl 7300} \\

 154 &  6513 & $\pm$&977  & {\sl  5404} & & & 154 & 8328 & 7242&$\pm$&1130 & {\sl 7213} \\

     &       &      &     &             & & & 178 & 7557 & 6543&$\pm$&778  & {\sl 6615} \\

 240 &  3821 & $\pm$&573  & {\sl  3964} & & & 240 & 6186 & 4855&$\pm$&1170 & {\sl 5525} \\

 408 &  2652 & $\pm$&325  & {\sl  2694} & & & 408 & 4335 & 4565&$\pm$&636  & {\sl 3993} \\

 614 &  1809 & $\pm$&127  & {\sl  1973} & & & 614 & 3297 & 3149&$\pm$&319  & {\sl 3088} \\

 750 &  1680 & $\pm$&100  & {\sl  1686} \\

1400 &  1016 & $\pm$&47   & {\sl  1011} & & & 1407 & 1890 & 1860&$\pm$&212 & {\sl 1781} \\

     &       &      &     &             & & & 1465 & 1840 & 1770&$\pm$&57  & {\sl 1732} \\

     &       &      &     &             & & & 1490 & 1820 & 1683&$\pm$&53  & {\sl 1712} \\

     &       &      &     &             & & & 1665 & 1690 & 1500&$\pm$&64  & {\sl 1583} \\ 

2695 &   571 & $\pm$&30   & {\sl   570} & & & 2695 & 1224 & 1176&$\pm$&57  & {\sl 1114} \\

4850 &   339 & $\pm$&25   & {\sl   332} & & & 4860 &  805 &  748&$\pm$&55  & {\sl  704} \\

4860 &   347 & $\pm$&17   & {\sl   331} & & & 4860 &  805 &  685&$\pm$&50  & {\sl  704} \\

     &       &      &     &             & & & 4995 &  795 &  679&$\pm$&50  & {\sl  689} \\

10700 &  176 & $\pm$&30   & {\sl   154} & & & 8085 &  525 &  505&$\pm$&55  & {\sl  462} \\

14900 &  112 & $\pm$&20   & {\sl   111} & & &14940 &  300 &  267&$\pm$&43  & {\sl  270} \\

      &      &      &     &             & & &15000 &  300 &  250&$\pm$&80  & {\sl  269} \\

      &      &      &     &             & & &22460 &  200 &  184&$\pm$&30  & {\sl  186} \\

      \\

\hline

\end{tabular*} 

\end{center}

}

\end{table}

\section{Constraints from the X-ray observations}

\subsection{Hot gaseous environment}

The radio galaxy 3C\,293 was recently observed with the {\sl Chandra} Advanced CCD Imaging Spectrometer (ACIS). The data collected on 2010 November were used by Lanz et al. (2015) to model the X-ray spectra of the nucleus, the inner jets, and faint X-ray features extending along the outer radio lobes. In particular, Lanz et al. extracted the spectrum of the nuclear region in the circular $3.\arcsec 5$ aperture (radius of $\sim 3$\,kpc) centred at the radio core precisely determined with the MERLIN/VLA/VLBI observations (Beswick, Pedlar \& Holloway 2002; Beswick et al. 2004) and discovered  that it could be fitted with a model consisting of a thermal component with a temperature of $1.03 \pm 0.10$\,keV, and two power-law components: one absorbed ($N_{\rm H}=9.4\times 10^{22}$\,cm$^{-2}$) with the photon index of $1.32 \pm 0.36$, and the other one unabsorbed with the photon index of $0.57 \pm 0.87$. The authors associated the absorbed power-law component with the active nucleus, while the unabsorbed power law and the thermal components with an off-nuclear source and a diffuse emission around the nuclear region, respectively. Lanz et al. also fitted  the spectra extracted from the two smaller areas (labelled as NC and NE0 in our Figure\,3b); the resulting values of photon indices of $0.77 \pm 0.24$ and $0.67 \pm 0.20$, respectively, were consistent within the errors with the global model values.

\begin{table}[th!]


\scriptsize{

\caption{Model input parameters}

\begin{tabular*}{90mm}{lrrrr}

\hline

\hline

Parameter    & Symbol   & Outer lobes & Inner lobes  \\

\hspace{6mm}(1)&\hspace{-2mm} (2)      & (3)         & (4)          \\

\hline

\\

{\bf Observed:} \\

Angular size $(\times 2)$           & $LAS$               & 216$\arcsec$ & 4.3$\arcsec$ \\

Length $(\times 2)$                 & $D/\sin\theta$      & 247 kpc & 4.2 kpc \\

Aspect ratio                        & $AR$                & 3.0     & 2.0     \\

Luminosity spectral densities\\

at a number of observing frequencies\\ 

$i=1,2,3,...$           & $P_{\nu_i}$ & Note(1) & Note(2) \\

\\

{\bf Set:} \\

Adiabatic index of the lobes' material            & $\Gamma_{\rm lb}$   & 4/3   & 4/3 \\

Adiabatic index of the ambient medium           & $\Gamma_{\rm x}$    & 5/3   & 5/3 \\

Adiabatic index of the lobes' magnetic field           & $\Gamma_{\rm B}$    & 4/3   & 4/3 \\

Minimum electron Lorentz factor (injected)  & $\gamma_{\rm min}$  & 1     & 1   \\

Maximum electron Lorentz factor (injected)  & $\gamma_{\rm max}$  & 10$^{7}$ & 10$^{7}$ \\

Core radius of the ambient density distribution & $a_{0}$             & 2\,kpc & 2\,kpc \\

Slope of the ambient density distribution    & $\beta$             & 1.0   & 0.1 \\

Thermal particles within the lobes          & $k$                 & 0     & 0 \\

Jet viewing angle                 & $\theta$            & 50$\degr$ & 75$\degr$ \\

\\

\hline

\end{tabular*}

Notes (1) and (2): Relevant luminosities are calculated with flux densities given in column 2 and column 6 of Table 2, respectively.
}

\end{table}

\begin{table}[t!]

\scriptsize{

\caption{Model-free parameters}

\begin{tabular*}{90mm}{lccccccc}

\hline

\hline

{\bf Parameter}     &        \multicolumn{3}{c}{Outer lobes} & & \multicolumn{3}{c}{Inner lobes} \\

\hspace{7mm}(1)         &  (2)  &  (3)  & (4)&     &(5)&  (6)  & (7) \\

\hline

\\

{\bf Main:} \\

$\alpha_{\rm inj}$                     & 0.59&{\bf 0.61}& 0.63 && 0.56&{\bf 0.59}& 0.60 \\

$Q_{\rm j}$\,[$10^{36}$\,W]           & 1.6&{\bf 2.1}& 2.7 && 1.6 &{\bf 2.1} & 2.5 \\  

$\rho_{0}$\,[$10^{-23}$\,kg\,m$^{-3}$]& 4.4&{\bf 3.2}& 2.2 && 5.4 &{\bf 3.4} & 2.5 \\

$t_{\rm lb}$\,[Myr]                      &  75 &{\bf  62} &  50  && 0.4 &{\bf 0.3} & 0.3 \\

$\chi^{2}_{\rm red.}$                  &0.460&{\bf 0.429}&0.448&& 0.682&{\bf 0.430}& 0.490\\

\\

{\bf Derived:} \\

$\upsilon_{\rm h}/c$                   &0.004&{\bf 0.007}&0.007&&0.010&{\bf 0.013}& 0.015\\

$p_{\rm lb}$\,[$10^{-12}$\,N\,m$^{-2}$]&0.07&{\bf 0.07} & 0.08&& 45&{\bf 46} & 47 \\

$B$\,[nT]                                & 0.5 &{\bf 0.5}  & 0.5 && 12&{\bf 12} & 12 \\

$\cal{M}_{\rm lb}$                     & 1.2  &{\bf 1.4 }  & 1.8  && 5.8 &{\bf 7.4}  & 8.7  \\

\\

 \hline

\end{tabular*}

}

\end{table}

From the thermal component model fits, Lanz et al. (2015) derived the corresponding electron number density of $0.019$\,cm$^{-3}$, which gives the central mass density of the hot gas of $3.18\times 10^{-23}$\,kg\,m$^{-3}$, assuming a uniform distribution and spherical geometry. This mass density is therefore almost identical to that emerging from our dynamical modelling of the lobes using the DYNAGE algorithm (see the best-fit model values in Table\,4). Moreover, the gas temperature of $kT=1.0$\,keV following from the {\sl Chandra} data analysis, translates to the adiabatic sound speed of $c_{\rm s}\simeq 5.3\times 10^{5}$\,m\,s$^{-1}$\,$\simeq 0.0018\,c$, assuming the adiabatic index $\Gamma_{\rm x}$=5/3, and the mean atomic weight $\mu =0.62$. 
Using these values, the inner lobes' expansion velocity turns out to be highly supersonic, with the corresponding Mach number of $\mathcal{M}_{\rm lb} \simeq 7.5$. This, again, is consistent with the conclusion of Lanz et al. (2015) that the X-ray--emitting gas in UGC\,8782 has been shock-heated by expanding jets. In this context, we note that 3C\,293/UGC\,8782 is particularly rich in warm (100\,K) molecular hydrogen (see Ogle et al. 2010), and that the shocks induced by the supersonic expansion of the inner lobes in the system is the most likely explanation for heating of this gaseous component as well.

\begin{figure*}[t]

\centering

\includegraphics[width=180mm, angle=0]{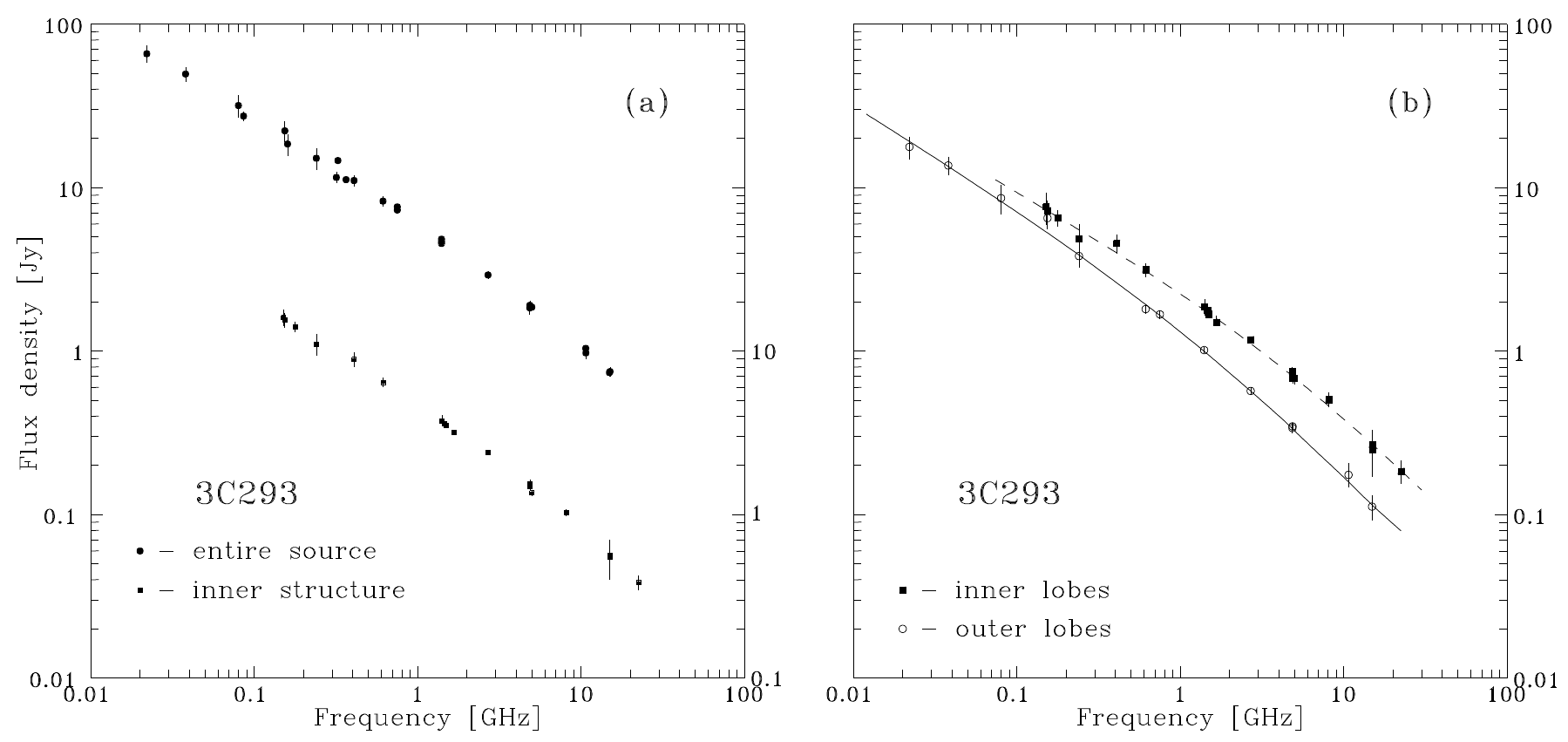}

\caption{(a) The observed broadband radio spectrum of the entire 3C\,293 radio source and of its inner structure; we note that the latter  is shifted one-decade down for a better visual inspection. (b) The estimated and modelled spectra of the outer and inner lobes. The dashed and solid curves indicate the best-fit models to the observational data points, which are given in columns (3) and (6) of Table~2.}

\end{figure*}

Lanz et al. (2015) also detected  several regions of the enhanced X-ray emission associated with the outer radio lobes. Two of them, labelled as NW2 and SE2, correspond to the outermost parts of the lobes, while the other two, NW1 and SE1, more or less  lie  inside the outer lobes (see the locations shown in our Figure\,2). All of these regions, except of the SE2, were fitted with both a power law and a thermal model. Only for the NW1 region was the thermal model  the best fit. Nevertheless, Lanz et al. could not rule out a possibility that all three regions were sources of a thermal emission from the hot gas expelled from the host galaxy and compressed by the expanding jets. If so, the emerging gas temperature of $\sim 3.7$\,keV would lead to the sound velocity of $c_{\rm s}\simeq 1.0\times 10^{6}$\,m\,s$^{-1}$\,$\simeq 0.0034\,c$, and hence only a mildly supersonic expansion of the outer lobes with the corresponding Mach number of $\mathcal{M}_{\rm lb} \simeq 1.5$.

\subsection{The accretion power}

Spherical (Bondi) accretion rate of the X-ray emitting gas onto the black hole in 3C\,293
\begin{equation}
\dot{M}_{\rm B}\approx \pi\, c_{\rm s}\, r_{\rm A}^{2}\, \rho_{\rm A},
\end{equation}
can be found knowing the sound speed $c_{\rm s}$ in the gaseous environment, the accretion radius $r_{\rm A}= 2 G M_{\rm BH}/c_{\rm s}^{2}$ where $M_{\rm BH}$ is the black hole mass, and the gas density at the accretion radius, $\rho_{\rm A}$. The black hole mass hosted by the galaxy UGC\,8782 was estimated from the $\log M_{\rm BH}-\log\sigma_{*}$ relation of Graham et al. (2011) with $\sigma_{*}=205.4\pm 7.3$\,km\,s$^{-1}$ given in the SDSS catalogue, and independently by Singh et al. (2015) from the $\log M_{\rm BH}-\log L_{\rm bulge}$ relation; the corresponding values of $1.55\times 10^{8}M_{\odot}$ and $(1.7\pm0.5) \times 10^{8}M_{\odot}$, respectively, are in a good agreement. Hence, in all the calculations below, we take $M_{\rm BH}/M_{\odot}=1.6\times 10^{8}$. This, along with the sound speed evaluated in the previous section as $c_{\rm s} \simeq 0.0018\,c$, gives $r_{\rm A}\simeq 5$\,pc.

To estimate the value of $\rho_{\rm A}$, we anticipate a significant increase in the gas density between $r \sim 3$\,kpc radius resolved with {\sl Chandra}, and $r_{\rm A}$. Following Allen et al. (2006), who analysed the X-ray emission of nine bright elliptical NGC galaxies, and, in particular, we assume a power-law scaling $\rho_{\rm A}=\rho(r)(r/r_{\rm A})^{m}$ with $m$ varying from $0.5$ to $1.0$. For $m=0.75\pm 0.25$, one therefore has $\rho_{\rm A}\simeq (4^{+16}_{-3})\times 10^{-21}$\,kg\,m$^{-3}$, and the Bondi accretion rate $\dot{M}_{\rm B} \simeq (1.8^{+7}_{-1.4})\times 10^{20}$\,kg\,s$^{-1}$ $\sim (3^{+11}_{-2.4}) \times 10^{-3}$\,M$_{\odot}$\,yr$^{-1}$. In the Eddington units, this rate corresponds to a range $\dot{M}_{\rm B}/\dot{M}_{\rm Edd} \sim (0.2-4) \times 10^{-2}$. We note that, with the standard $10\%$ radiative efficiency of the accretion disk, the implied bolometric AGN luminosity would then read  $0.1 \dot{M}_{\rm B} c^2\, \sim (1.6^{+6.3}_{-1.2})  \times 10^{36}$\,W. Meanwhile, the actual bolometric AGN power in the system, estimated from the observed H$\alpha$ luminosity of $L_{\rm H \alpha} \simeq 1.5 \times 10^7 \, L_{\odot}$ (Sikora et al. 2013), is much higher than this, namely

\begin{equation}
L_{\rm bol} \simeq 2.0 \times 10^3 \, L_{\rm H \alpha}  \sim 1.2 \times 10^{37}\,{\rm W} \, .
\end{equation}
Hence, we conclude that the accretion rate in 3C\,293/UGC\,8782 is relatively high, $\dot{M}_{\rm acc}/\dot{M}_{\rm Edd} \sim 0.1$, exceeding the spherical Bondi accretion rate of a hot gas. This should not be of any surprise, however, since the analysed system is rich in warm and cold molecular gas, with the total masses of $3.7 \times 10^9 M_{\odot}$ and $2.2 \times 10^{10} M_{\odot}$, respectively (Ogle et al. 2010; Labiano et al. 2014). Importantly, this gas is distributed along the extended warped disk, which rotates around the galactic centre.

\section{Discussion}

The main clues for understanding the complex radio morphology of 3C\,293 are (i) the presence of two optical nuclei in the centre of the host galaxy, which are separated by $\sim 880$\,pc, (ii) the radio spectra of the outer lobes, which  indicates a very recent termination in a supply of fresh electrons, and that this also coincided with the formation of young ($\sim 0.3$\,Myr) and misaligned inner lobes in the system, and (iii) the same kinetic power of the jets supplying the outer and inner lobes.

\subsection{The jet production efficiency}

For the given accretion power estimated in the previous section based on the H$\alpha$ luminosity, $\dot{M}_{\rm acc} c^2 \simeq 10^{38}$\,W, the jet kinetic luminosity emerging from our modelling, $Q_{\rm j} \simeq 2 \times 10^{36}$\,W, can be compared to the power of a Blandford-Znajek outflow (Blandford \& Znajek 1977). The maximum efficiency in the production of this, $P_{\rm BZ}$, is enabled in the case of the magnetically arrested accretion scenario, where the amount of the magnetic flux around the black hole is large enough to affect the accretion process itself (e.g., Tchekhovskoy et al. 2010). With the dimensionless angular momentum (spin) of the black hole, $a=c J_{\rm BH}/G\,M_{\rm BH}^{2}$,  in particular one finds

\begin{equation}
P_{\rm BZ} \simeq 10 \, \left(\phi_{\rm BH}/50\right)^{2} \, x^{2}_{\rm a} \, f\!(x_{\rm a}) \,\, \dot{M}_{\rm acc} c^{2} \, ,
\end{equation}
where $x_{\rm a}=a/2 [1+\sqrt{1-a^{2}}]$, $f(x_{\rm a})\simeq 1+1.4\,x^{2}_{\rm a}-9.2\,x^{4}_{\rm a}$, and $\phi_{\rm BH}$ is the dimensionless magnetic flux (see e.g., Sikora et al. 2013). Hence, the condition $Q_{\rm j} \sim P_{\rm BZ}$ is satisfied for relatively low value of the black hole spin, namely $a \sim 0.2$. Thus, we conclude that either the supermassive black hole in the 3C\,293/UGC\,8782 is indeed accreting in the magnetically arrested regime, but the black hole spin value is low, or that the black hole spin is high (maximum) but instead the magnetisation of the accretion disk is low. We will come back to this issue in Section 5.3, below.

\subsection{X-shaped radio morphology}

As noted in the Introduction, overall radio morphology of 3C\,293 resembles an X-shaped source, in which one pair of lobes is oriented along a substantially different axis with respect to the other one, leading to the formation of an X-like structure (Leahy \& Parma 1992). The inner lobes of 3C\,293 are, however, two orders of magnitude shorter than the outer ones. Despite the fact that our knowledge of these peculiar group of sources is growing continuously, and the number of such objects known is increasing (Cheung 2007), the physical cause of this type of morphological arrangement remains unclear. Several authors have attempted to explain the formation of X-shaped structures and a comprehensive review of the existing models is given by Gopal-Krishna et al. (2012). The scenarios proposed include: (i) an anisotropic hydrodynamic backflow (Leahy \& Williams 1984; Worrall, Birkinshaw \& Cameron 1995; Hodges-Kluck \& Reynolds 2011), (ii) an interaction of the jet with merger remnants within the host galaxy (Gopal-Krishna et al. 2012), (iii) a binary black hole system producing two pairs of jets (Lal \& Rao 2007), (iv) a conical precession of the jet (Parma, Ekers \& Fanti 1985), or finally (v) a fast realignment of the jet axis (Dennett-Thorpe et al. 2002; Merritt \& Ekers 2002).

In the particular case of 3C\,293, the fact that the misaligned pair of compact lobes emerges from the galactic centre, invalidates the hydrodynamic backflow/interaction models (i--ii). The binary black hole hypothesis of Lal \& Rao (2007), on the other hand, sounds more convincing owing to the presence of two optical nuclei (as expected in a post-merger system; see, e.g., Colpi \& Dotti 2011). We note in this context that some close binary active nuclei with multiple radio jets are known, including 3C\,75 with the core separation of 8\,kpc (e.g., Hudson et al. 2006), PKS\,2149$-$158, which is a triple radio galaxy with the core separation of about 16\,kpc (Guidetti et al. 2008), or 3C\,338 (Ge \& Owen 1994). Another interesting example is the quasar J1502$+$1115 that contains a close-pair binary (separated by 138\,pc), and a third radio component that is located about 7\,kpc away (Deane, et al. 2014). Yet the fact that the jet kinetic power in 3C\,293 is the same for both the inner and the outer pair of lobes, as revealed by our modelling, would be rather difficult to explain in the framework of the model (iii), taking into account that masses and spins of the two black holes are, most likely, significantly different. The jet precession scenario (iv) is also, to some extent, supported by the wiggling appearance of the inner lobes with the tail-like extension bending out of the galactic disk. Still, the radio spectral and morphological properties of the inner and outer lobes in 3C\,293 indicate these are quite distinct episodes of the jet activity. In other words, while the inner jets in the system may indeed be precessing, we consider the jet precession  an unlikely cause of a dramatic misalignment of the outer and inner radio structures. All in all, we conclude that a rapid realignment of the jet axis model (v) is the most attractive one.

A fast realignment of the jet axis may have resulted from a rapid flip of the black hole spin,  either because of a black hole-black hole merger, or a tilted accretion disk. Again, the presence of two optical nuclei in the studied object favour the latter option. And indeed, 3C\,293/UGC\,8782 is a post-merger system, in which plentiful cold and warm molecular gas supplied by the spiral companion is distributed along the extended warped disk. Bearing in mind a complex kinematics of a merger process, a misalignment between the angular momentum of this gaseous structure and of a primary black hole spin can be expected. In this situation, matter accreting continuously onto the black hole would then flip the black hole spin on the timescale of 

\begin{equation}
\frac{\tau_{\rm align}}{{\rm Myr}} \sim 0.3 \left(\frac{a}{0.1}\right)^{11/16} \left(\frac{\alpha_{\rm disk}}{0.03}\right)^{13/8} \left(\frac{M_{\rm BH}}{10^8 M_{\odot}}\right)^{-1/16} \left(\frac{\dot{M}_{\rm acc}}{0.1 \dot{M}_{\rm Edd}}\right)^{-7/8}
\end{equation}
where $\alpha_{\rm disk}$ is the Shakura-Sunyaev accretion disk viscosity parameter (Natarajan \& Pringle 1998). This, for the conditions advocated here for 3C\,293, namely $a \sim 0.2$, $\dot{M}_{\rm acc} \sim 0.1\, \dot{M}_{\rm Edd}$, and $M_{\rm BH} \simeq 1.6 \times 10^8 M_{\odot} $, and assuming the widely anticipated $\alpha_{\rm disk} \sim 0.03$, is exactly of the order of the sound-crossing timescale within the outer radio lobes, $\tau_{\rm s/lb}$ (see Section\,3.2), assuring a self-consistency in our interpretation.

\subsection{X-shaped radio galaxies: low values of the black hole spins?}

In Section\,5.1 above, we concluded that the supermassive black hole in 3C\,293/UGC\,8782 is either accreting in the so-called magnetically arrested regime but the black hole spin value is low ($a\sim 0.2$), or that the black hole spin is high, but instead the magnetisation of the accretion disk is low. Here we posit the hypothesis that the former option is correct, and moreover that it applies to other X-shape radio galaxies as well. That is, we propose that the X-shape radio morphology forms in post-merger systems that are rich in cold molecular gas, which assures an efficient accretion (see Mezcua et al. 2012), with large net magnetic flux accumulated in central regions of the accretion flow that only host, however,  slowly spinning supermassive black holes (likely in close binary systems with others, less massive black holes; Liu 2004). Tilted accretion disks naturally formed in these systems, along with the postulated low values of black hole spins, enable a rapid realignment of the jet axis, leading to the formation of winged radio morphologies. This scenario would then also explain relatively low radio luminosities of X-shaped radio galaxies in general, intermediate in-between Fanaroff-Riley type I and type II sources (Cheung 2007).

\begin{acknowledgements}
Authors acknowledge the referee for very useful suggestions that enabled us to improve this paper. 
J.M., M.J., and M.W. were supported by the Polish NSC grant DEC-2013/09/B/ST9/00599. {\L}.S. was supported by Polish NSC grant DEC-2012/04/A/ST9/00083.

\end{acknowledgements}

\end{document}